\documentclass[a4paper,10pt,twoside]{cpc-hepnp}

\usepackage{multicol}
\usepackage{graphicx}
\usepackage{booktabs}
\usepackage{amssymb,bm,mathrsfs,bbm,amscd}
\usepackage[tbtags]{amsmath}
\usepackage{lastpage}

\begin{document}

\fancyhead[c]{\small Submitted to Chinese Physics C}
\fancyfoot[C]{\small \thepage}


\title{Identification of a $9/2^-$[505] isomer in the neutron-rich $^{193}$Os nucleus\thanks{Supported by National Natural Science
Foundation of China (10905075) }}

\author{%
\quad B. S. Gao$^{1,2,3}$
\quad X. H. Zhou$^{1;1)}$\email{zxh@impcas.ac.cn}%
\quad Y. D. Fang$^{1}$
\quad Y. H. Zhang$^{1}$
\quad M. L. Liu$^{1}$
\quad S. C. Wang$^{1,2}$
\quad J. G. Wang$^{1}$\\
\quad F. Ma$^{1}$
\quad Y. X. Guo$^{1}$
\quad X. G. Wu$^{4}$
\quad C. Y. He$^{4}$
\quad Y. Zheng$^{4}$
\quad Z. M. Wang$^{4}$\\
\quad X. L. Yan$^{1,2}$
\quad Z. G. Wang$^{1,2,3}$
\quad F. Fang$^{1}$
}
\maketitle

\address{%
$^1$ Institute of Modern Physics, Chinese Academy of Sciences, Lanzhou 730000, People's Republic of China\\
$^2$ Graduate University of Chinese Academy of Sciences, Beijing, 000049, People's Republic of China\\
$^3$ School of Nuclear Science and Technology, Lanzhou University, Lanzhou 730000, People's Republic of China\\
$^4$ China Institute of Atomic Energy, Beijing 102413, People's Republic of China\\
}

\begin{abstract}
The neutron rich nucleus $^{193}$Os was produced in the $^{192}$Os($^{7}$Li,$^{6}$Li)$^{193}$Os reaction.
An isomeric state based on the $9/2^-$[505] nilsson orbital was identified in the present work.
Half-life of the isomeric state was extracted and discussed in terms of the $K$ quantum number.
Level scheme built on the isomeric state was proposed based on the experimental data.

\end{abstract}

\begin{keyword}
high-spin state, isomeric state, $K$-forbidden, hindrance factor
\end{keyword}

\begin{pacs}
21.10.Tg,23.35.+g,23.20.Lv
\end{pacs}


\begin{multicols}{2}

\section{Introduction}

The nucleus $^{193}$Os  lies at the neutron-rich side of the stability valley and thus cannot easily be produced in fusion-evaporation reactions with stable beam and target combinations.
Spectroscopic information of this nucleus to date comes from neutron capture \cite{ZPhysA.285.405, NuclPhysA.316.13, FizikaB.11.83} and $(d,p)$ transfer \cite{ZPhysA.285.405} reactions, which involve small angular momentum transfers.
As a result, only low-spin levels in the nucleus $^{193}$Os were observed in the previous work.
In lighter odd-$A$ osmium isotopes $^{185-191}$Os, high-spin level structures built on high-$j$ Nilsson configurations such as 11/2$^+$[615] were systematically established \cite{NuclPhysA.237.333, PhysRevC.14.2095, JPhysG.31.S1891}.
On the other hand, high-spin band structures built on the 13/2$^+$[606] configuration were observed in heavier $^{195,197}$Os nuclei via the fragmentation of an $E/A=1$ GeV $^{208}$Pb beam \cite{PhysRevC.84.044313}.
Thus in the odd-$A$ Os isotopes up to mass number $A=197$, high-spin states have been experimentally investigated in the previous work except for the nucleus $^{193}$Os.
During the course of this investigation, S.~J.~Steer $et$ $al.$ reported an isomeric state in $^{193}$Os with a half-live of $T_{1/2}=132(29)$ ns through the fragmentation of a $^{208}$Pb beam \cite{PhysRevC.84.044313}.
However, due to low statistics, they were not able to get detailed information of the isomeric state, such as excitation energy, spin-parity and configuration.
In this article, we present the new results of the isomeric state in $^{193}$Os, which were obtained as a by-product in the $^{7}$Li + $^{192}$Os experiment.

\section{EXPERIMENT AND RESULTS}
\subsection{Measurements}
The experiment was aimed to investigate the high-spin states of $^{194,195}$Au through fusion-evaporation $^{192}$Os($^{7}$Li, $xn\gamma$)$^{194,195}$Au($x=5,4$) reactions \cite{PhysRevC.86.054310, PhysRevC.85.027301}.
The $^{7}$Li beam was provided by the HI-13 Tandem Accelerator at China Institute of Atomic Energy in Beijing (CIAE).
The target was an isotopically enriched $^{192}$Os metallic foil of 1.7 mg/cm$^{2}$ thickness with a 1.1 mg/cm$^{2}$ carbon backing to stop the recoiling nuclei.
The $^{193}$Os nucleus was produced via the one neutron-transfer $^{192}$Os($^{7}$Li,$^{6}$Li)$^{193}$Os reaction.
$X$-$\gamma$-$t$ and $\gamma$-$\gamma$-$t$ coincidence measurements were performed at a beam energy of 44 MeV.
Here, $X$ refers to x rays and $t$ refers to the relative time difference between the two coincident $\gamma$ rays of at most 400 ns in our experiment.
An array of 14 Compton-suppressed HPGe detectors was used to detect the $\gamma$ rays emitted.
The energy and efficiency calibrations were made using $^{60}$Co, $^{133}$Ba, and $^{152}$Eu standard sources.
The systematic errors for the energies of $\gamma$ rays were estimated to be 0.1 $\sim$ 0.6 keV depending on energy region.
Energy resolutions of the Ge detectors were about 2.0 $\sim$ 2.5 keV at full width at half maximum for the 1332.5-keV line.
A total of 9$\times$10$^{7}$ coincidence events were accumulated.
After accurate gain matching, the $\gamma$-$\gamma$ coincidence data were sorted off-line according to the energies of the two $\gamma$ rays into two $4K\times4K$ matrices under two different coincidence time conditions.
The $\gamma$-ray peak, which was the start signal in a time-to-amplitude converter(TAC), was used to make gated spectra, and the coincidence spectra were then grouped according to two coincidence time ranges: (1) $t<$50 ns, defined here as prompt coincidences; (2) 150 ns $<t<$ 400 ns, as delayed coincidences.
The half-life of the isomeric state was also extracted from the $\gamma$-$\gamma$-$t$ coincidence data.

In the experiment carried out at the Laboratori Nazionali di Legnaro, Italy, $^{193}$Os was produced in the $^{82}$Se + $^{192}$Os collision system at a bombarding energy of $E$($^{82}$Se) = 460 MeV.
Three-fold coincidence data was collected in the experiment.
As a cross-check of the results from the $^{7}$Li+$^{192}$Os reaction, the $\gamma$-ray coincidence relationships in the $^{82}$Se + $^{192}$Os reaction were analyzed as well.
More detailed description of this experimental setup can be found in Ref. \cite{PhysRevC.70.024301}.

\subsection{Identification of the isomeric state}
The isomeric state was observed from the analysis of $\gamma$-ray coincidence relationships in $^{194}$Au where there exists a 243.6-keV transition\cite{PhysRevC.86.054310}.
We produced two spectra gated on 242.7-keV transition showing $\gamma$ rays in prompt and delayed coincidence with the 242.7-keV transition (see Fig.~\ref{Gated_242}).
We can see from panel (a) in Fig.~\ref{Gated_242} that the 242.7-keV transition is in prompt coincidence with itself as well as $\gamma$ rays from $^{194}$Au and Au x-rays.
However, the delayed coincidence relationships presented in panel (b) shows that the 242.7-keV transition is in delayed coincidence with only itself.
None of the $\gamma$ rays from $^{194}$Au is in delayed coincidence with the 242.7-keV transition.
This suggests that there exists two $\gamma$ rays with energy around 242.7-keV in a nucleus, and one of them feeds an isomeric state and the other one deexcites the isomeric state.
A close inspection of panels (a) and (b) reveals that the 242.7-keV transitions are in coincidence with osmium x-rays as well as a 72.9-keV $\gamma$ ray.
Thus we assigned the two 242.7-keV lines to osmium through coincidences with characteristic x rays.
In order to assign the two 242.7-keV $\gamma$ rays to a specific Os nucleus, we systematically studied low-spin states data for Os isotopes.
We suggest that the 242.7-242.7-keV cascade belongs to the $^{193}$Os nucleus because a 72.9-keV transition exists in the low-lying levels in $^{193}$Os \cite{NuclPhysA.316.13}.
This assignment of the 242.7-242.7-keV cascade is further supported by a recent work in which an isomeric state depopulated by a 242-keV transition is discovered in $^{193}$Os \cite{PhysRevC.84.044313}.
In order to extract the half-life of the isomeric state, a least square fitting to the decay curve of the 242.7-keV $\gamma$ ray is performed, and the result is presented in fig.~\ref{decay_slop}.
Our half-life is $T_{1/2} = 110(28)$ ns, similar to the value of $T_{1/2} = 132(29)$ ns determined previously \cite{PhysRevC.84.044313}.

By analyzing $\gamma$-ray coincidence relationships in the $^{82}$Se + $^{192}$Os reaction, four more transitions were discovered above the 242.7-242.7-keV cascade, namely 309.4-keV, 328.2-keV, 412.2-keV and 346.5-keV transitions (see Fig.~\ref{LevelScheme}).
Representative coincidence spectra produced using double gates on the data collected in the $^{82}$Se+$^{192}$Os experiment are given in Fig.~\ref{DoubleGates}.
The assignments of the new $\gamma$ rays, as indicated in Fig.~\ref{LevelScheme}, to $^{193}$Os are all checked by the cross $\gamma$-ray coincidences (the $\gamma$ rays coming from the decay of the ``target-like'' fragments in coincidence with those coming from the ``beam-like'' reaction products); the 242.7-, 309.4-, 328.2-, 412.2-, and 346.5-keV lines are found to be in strong coincidence with the 9/2$^+$ $\rightarrow$ 7/2$^+$, 191-keV transition in $^{81}$Se \cite{EPJA.39.295} (see Fig.~\ref{DoubleGates}).
The level scheme for $^{193}$Os established in the present work is shown in Fig.~\ref{LevelScheme}.

The measured $\gamma$-ray energies, relative intensities, ADO ratios and suggested spin and parity assignments are summarized in Table \ref{ADOs}.
ADO ratios for the two 242.7-keV transitions are not available since the two transitions cannot be distinguished in gated spectra. 
Spin and parity assignments for most of the new levels cannot be uniquely determined due to insufficient experimental information.

\begin{center}
\includegraphics[height=6cm]{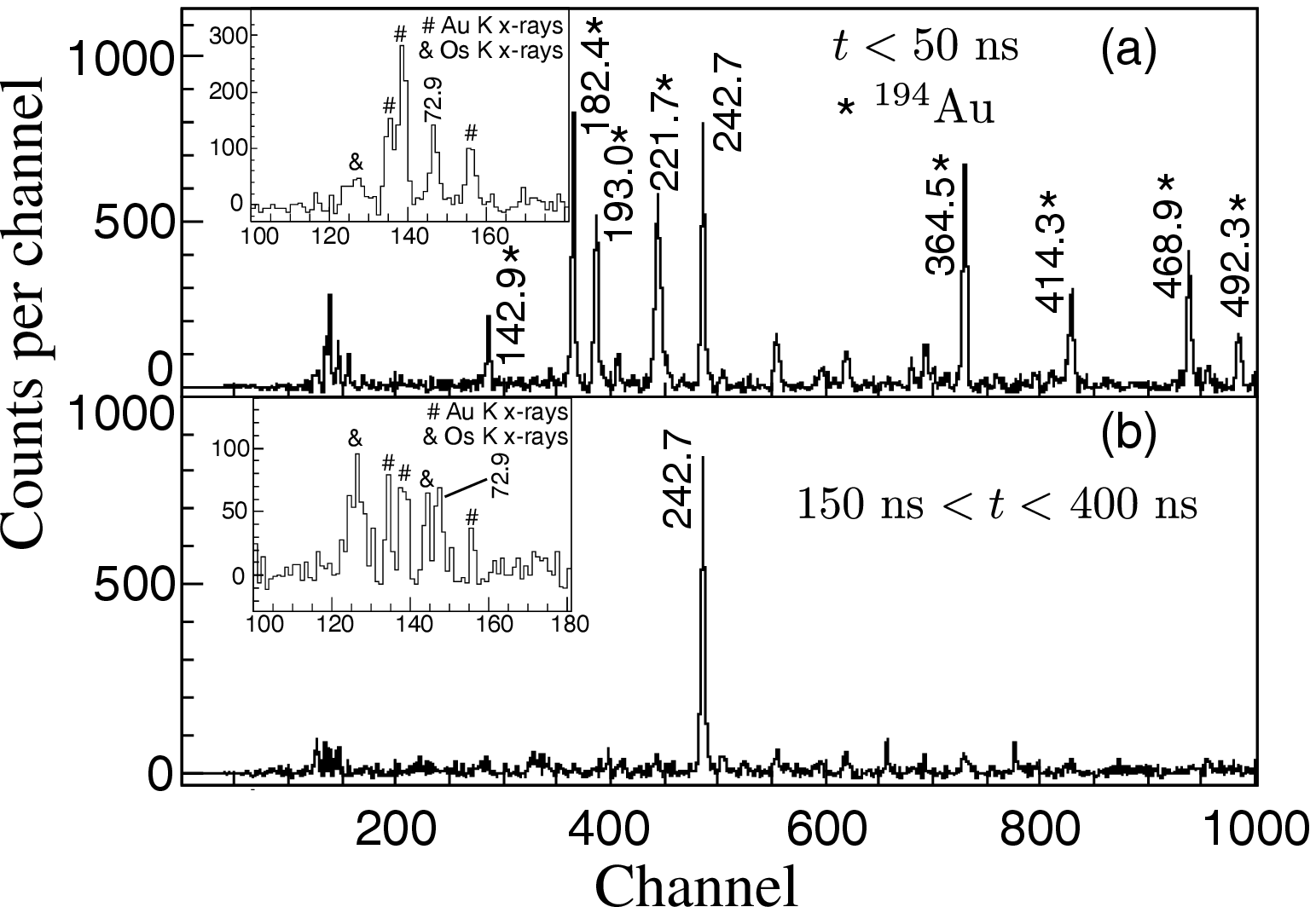}
\figcaption{\label{Gated_242} Spectra gated on the 242.7-keV transition showing $\gamma$-rays in prompt [panel (a)]  and delayed [panel (b)] coincidence with the 242.7-keV transition. Contaminations from $^{194}$Au are marked with asterisks.}
\end{center}
\begin{center}
\includegraphics[height=8cm]{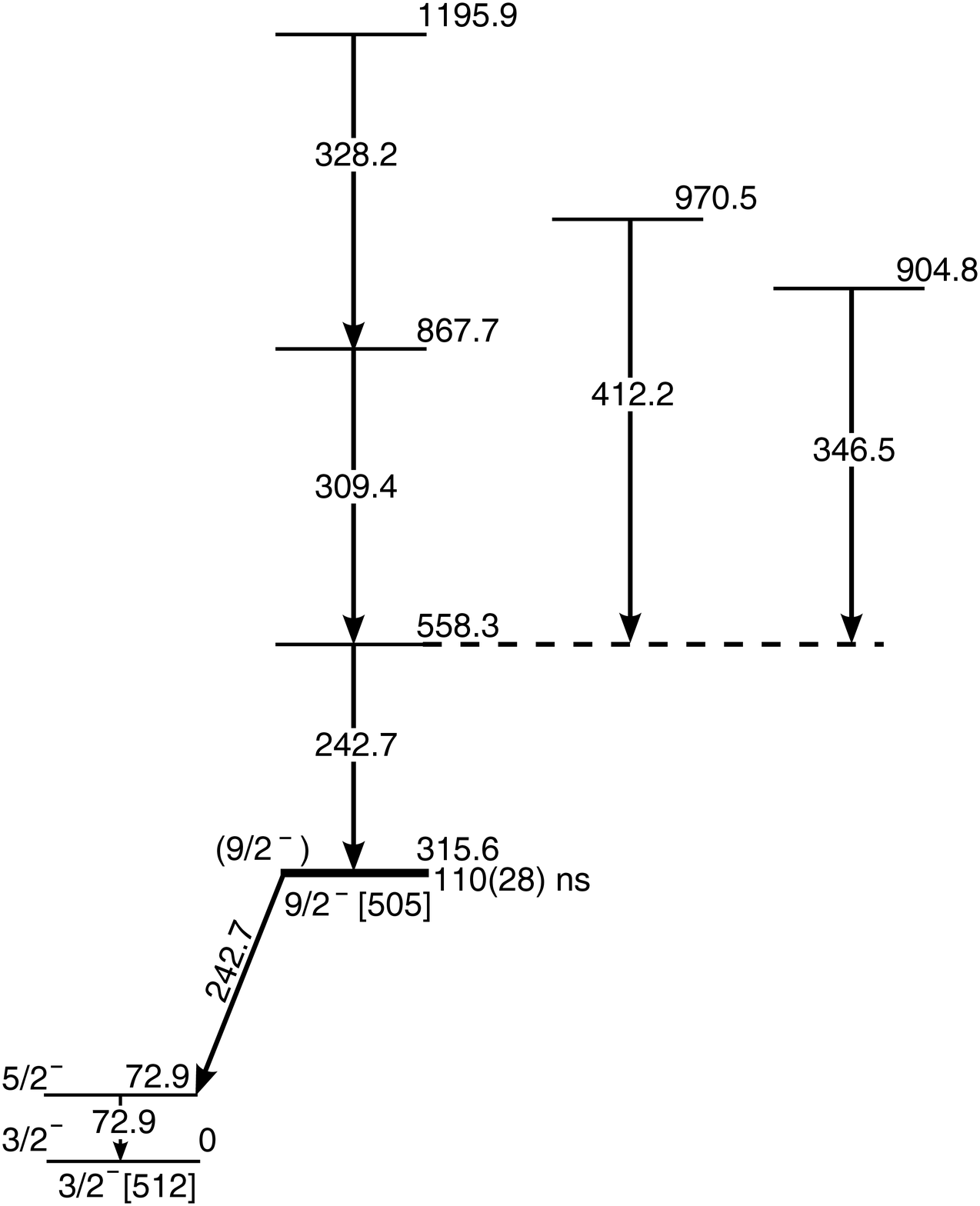}
\figcaption{\label{LevelScheme} Level scheme for $^{193}$Os deduce from the present work.}
\end{center}

\begin{center}
\includegraphics[height=6cm]{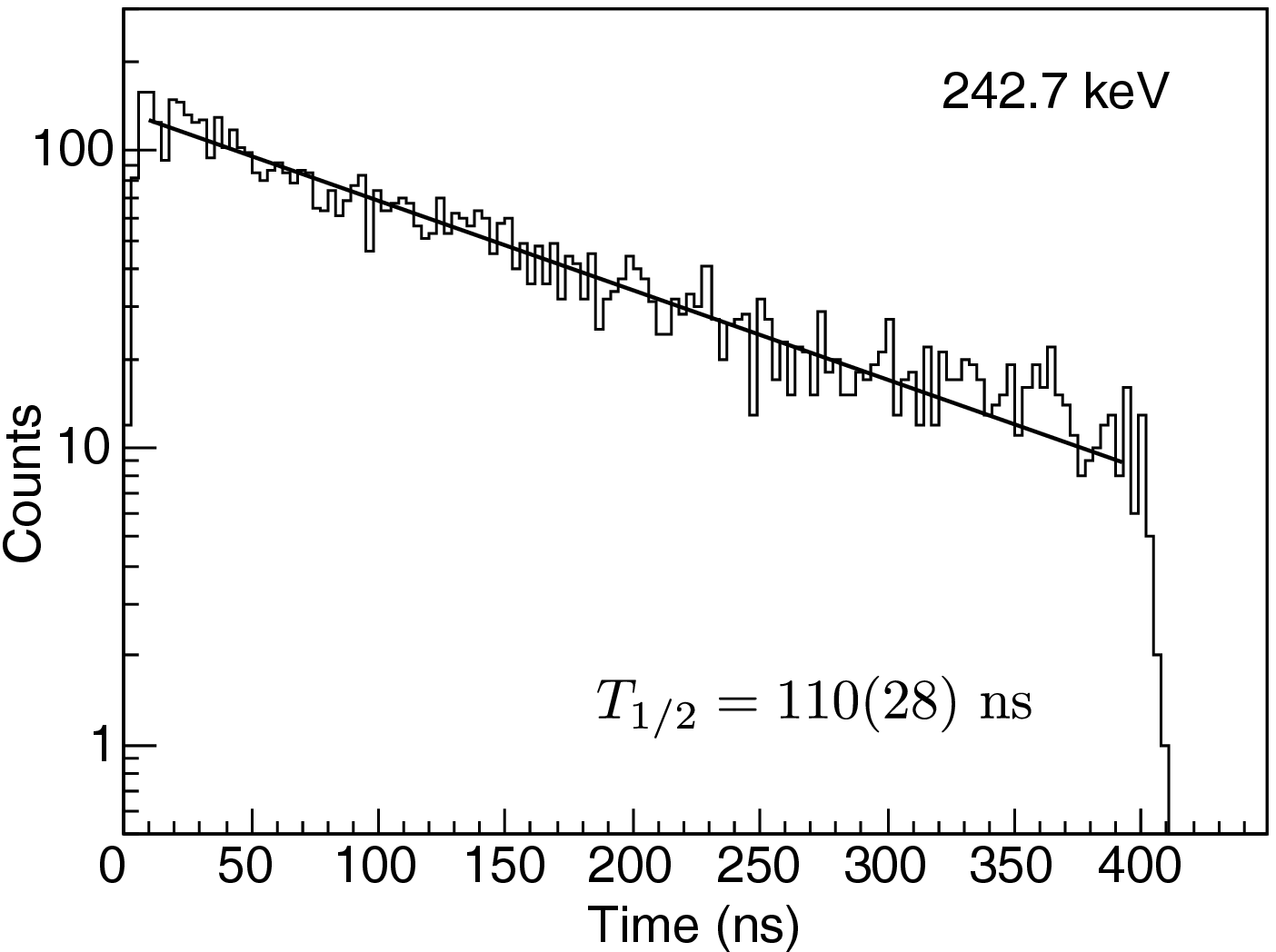}
\figcaption{\label{decay_slop} Least square fitting to the decay curve of the 242.7-keV transition.}
\end{center}

\begin{center}
\includegraphics[height=8cm]{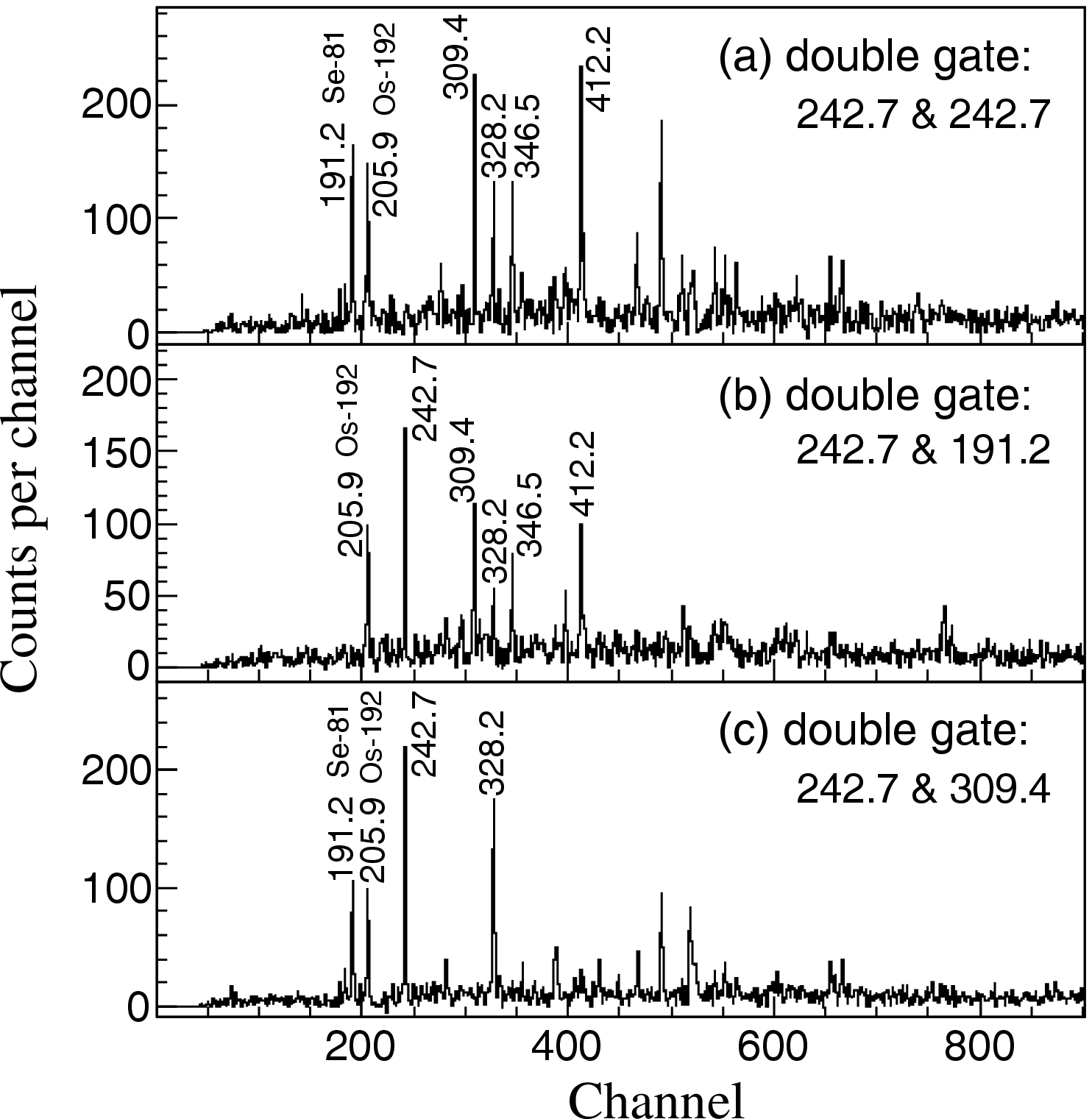}
\figcaption{\label{DoubleGates} $\gamma$-ray coincidence spectra produced using double gates on the (a) 242.7- and 242.7-keV lines, (b) 242.7- and 191.2-keV lines, (c) 242.7- and 309.4-keV lines. The 205.9- and 191.2-keV transitions are from the target nucleus $^{192}$Os (random coincidence) and the partner nucleus $^{81}$Se (see text), respectively.}
\end{center}
\begin{center}
\tabcaption{ \label{ADOs}  $\gamma$-ray transition energies, relative intensities, ADO ratios, and their assignments in $^{193}$Os.}
\footnotesize
\begin{tabular*}{80mm}{c@{\extracolsep{\fill}}cccc}
\toprule $E_\gamma$(keV) & $I_\gamma^a$  & $R_{\rm ADO}^a$  & $E_i \rightarrow E_f$(keV) & $J_i^\pi \rightarrow J_f^\pi$  \\
\hline
73.2  &         &            & 73.2 $\rightarrow$ 0    &  5/2$^-$ $\rightarrow$ 3/2$^-$ \\
242.7 &         &            & 315.9$\rightarrow$ 73.2 & (9/2$^-$)$\rightarrow$ 5/2$^-$ \\
242.7 & 100(15) &            & 558.6$\rightarrow$315.9 &                    \\
309.4 & 27(4)   & 0.74(10)   & 868.0$\rightarrow$558.6 &                    \\
328.2 & 17(3)   & 0.74(13)   &1196.2$\rightarrow$868.0 &                    \\
412.2 & 40(6)   & 1.63(15)   & 970.8$\rightarrow$558.6 &                    \\
346.5 & 19(3)   & 0.63(10)   & 905.1$\rightarrow$558.6 &                    \\
\bottomrule
\end{tabular*}
$^a$Extracted in the $^{82}$Se+$^{192}$Os experiment.~~~~~~~~~~~~~~~~~~
\vspace{0mm}
\end{center}

\section{DISCUSSION}

($^{7}$Li,$^{6}$Li) reactions involve the transfer of a neutron that is fairly tightly bound in the projectile ($S_n = 7.25$ MeV). 
Thus, the ground-state reaction $Q$ values are generally negative and populations of high-spin states at low excitation energies are favored.\cite{PhysRevC.16.1456}
The $Q$ value of the $^{192}$Os($^{7}$Li, $^{6}$Li)$^{193}$Os reaction is -1.67MeV, which suggests that the isomeric state in $^{193}$Os should most probably have high angular momentum with relatively low excitation energy.
This indicates that this state is built on configurations associated with high-$j$ orbitals closed to the Fermi level.
A systematic investigation of the orbitals closest to the Fermi level (see Fig.~\ref{nilsson_level}) shows that the isomeric state should be associated with a neutron occupying one of the $13/2^+$[606], $11/2^+$[615] and $9/2^+$[624] orbitals in the $i_{13/2}$ subshell or the $9/2^-$[505] orbital in the $h_{9/2}$ subshell.
Assuming any of the above configurations for the isomeric state, the 242.7-keV $(9/2)^- \rightarrow 5/2^-$ transition is $K$-forbidden [electromagnetic transitions involving the $K$ change equal to or less than the transition multipolarity (i.e., $\Delta K \leqslant \lambda$) are allowed in the $K$ selection rule, otherwise the transitions ($\Delta K > \lambda$) are forbidden].
$K$-forbidden transitions can be discussed in terms of the hindrance factor $F=T_{1/2}^\gamma / T_{1/2}^W$ or the hindrance factor per degree of $K$ forbiddenness $f_\nu = F^{1/\nu}$, where $T_{1/2}^\gamma$ is the partial $\gamma$-ray half-life, $T_{1/2}^W$ is the corresponding Weisskopf single-particle estimate, and $\nu = \Delta K - \lambda$ is the order of $K$ forbiddenness.
Large number of $K$-forbidden transitions have been discovered in the rare earth region (with neutron numbers $89 \leqslant N \leqslant 114$ and proton numbers $62 \leqslant Z \leqslant 78$) and the actinide region ($N \geqslant 134$).
For nuclei in these regions, for each unit of $K$ change that exceeds the transition’s multipolarity the transition rate may be reduced by a factor of about one hundred \cite{PhysLettB.26.369}.
In the transitional W-Os-Pt region, hindrance factors for $K$-forbidden transitions have been studied for two nuclei, namely $^{187}$W \cite{PhysRevC.71.067301} and $^{191}$Os \cite{JPhysG.31.S1891}, adjacent to $^{193}$Os.
The value of $f_\nu$ were extracted to be $f_\nu <$ 100 and $f_\nu =$1.9 in $^{187}$W and $^{191}$Os, respectively.
Considering the arguments mentioned above, it's reasonable to conclude that the value of $f_\nu$ in $^{193}$Os might be in the range of several tens to several hundreds.
By assuming $f_\nu$ = 100 for the 315.6-keV isomeric state in $^{193}$Os, assignments of $13/2^+$[606], $11/2^+$[615], $9/2^+$[624] and $9/2^-$[505] configurations for this state would lead to half-lives of 24 year, 430 ms, 81 us and 540 ns, respectively.
We can see that, considering the half-life of the isomeric state, only the assignments of the $9/2^-$[505] or $9/2^+$[624] configurations are reasonable, as the other two configuration assignments would lead to much longer half-lives than the observed value of 110 ns.  
A systematic investigation of the excitation energies of the $9/2^+$[624] and $9/2^-$[505] configurations in Os isotopes is then performed.
It turns out that the isomeric state is more likely associated with the $9/2^-$[505] configuration.
The systematics of the Nilsson levels implies that with increasing odd neutron number at approximately constant deformation, each Nilsson orbital in turn will first be observed as a high lying particle excitation and ultimately become an increasingly deeper hole excitation.
We listed in TABLE \ref{excitation_energy} the excitation energies of the $9/2^+$[624] and $9/2^-$[505] configurations in odd-$A$ Os isotopes.
It can be seen that excitation energies of the two configurations follows well with the trend described above.
This implies that the $9/2^+$[624] configuration in $^{193}$Os would lie very high in energy and not likely to be observed in our experiment.
On the other hand, the $9/2^-$[505] configuration has been observed at excitation energies of 31 and 0 keV in $^{189}$Os and $^{191}$Os nuclei, respectively. 
This suggests that the $9/2^-$[505] configuration in $^{193}$Os lies low in energy as a hole excitation and might be experimentally observed.
In fact, the $9/2^-$[505] configuration is observed in our experiment at excitation energy of 315.6 keV.
The excitation energy of the $9/2^-$[505] configuration in odd-A Os isotopes fits well with the systematics, which supported our configuration assignment to the isomeric state.
Similar case was also observed in the neighboring nucleus $^{187}$W where the band head of the $9/2^-$[505] configuration decays to the $3/2^-$[512] band through $K$-forbidden $E$2 and $M$1 transitions \cite{NuclPhysA.619.1}.
According to the Weisskopf estimate with corrections for $K$ forbiddenness, half-life of the band head of the $9/2^-$[505] configuration in $^{187}$W was predicted to be in the range of about several ten ns \cite{NuclPhysA.619.1}, which is consistent with the case in $^{193}$Os.
\begin{center}
\includegraphics[height=9.5cm]{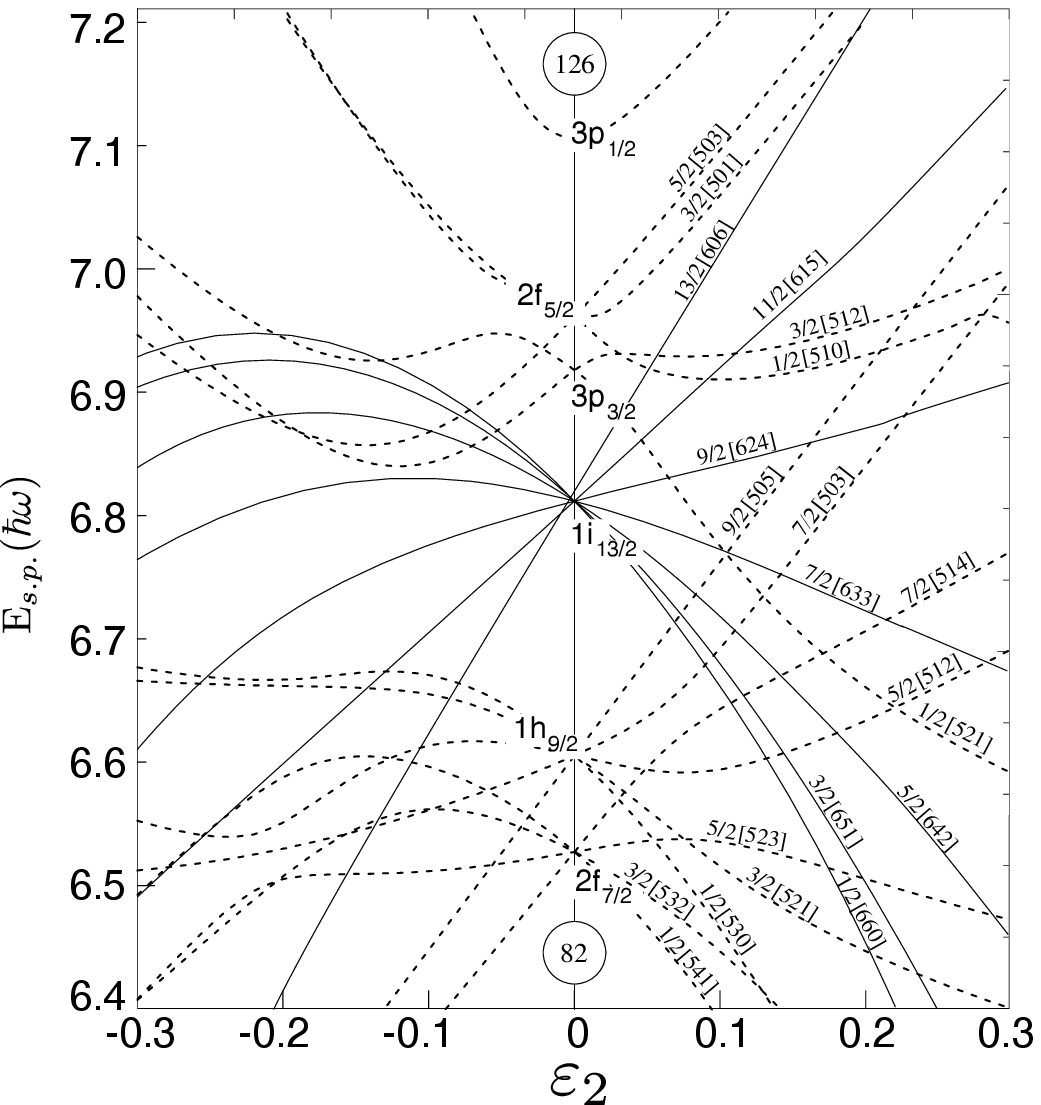}
\figcaption{\label{nilsson_level} Nilsson diagram for neutrons, taken from Ref.~\cite{NuclDataSheets.97.241}}
\end{center}

\begin{center}
\tabcaption{ \label{excitation_energy}  Excitation energies for the $9/2^+$[624] and $9/2^-$[505] configurations in odd-$A$ $^{181-193}$Os.}
\footnotesize
\begin{tabular*}{80mm}{c@{\extracolsep{\fill}}ccc}
\toprule Nuclei & $9/2^+$[624]  & $9/2^-$[505]  & Ref.  \\
\hline
$^{181}$Os & 157 & $-$ & \cite{NuclPhysA.728.287} \\
$^{183}$Os & 0   & $-$ & \cite{NuclPhysA.696.337} \\
$^{185}$Os & 402 & $-$ & \cite{NuclPhysA.237.333} \\
$^{187}$Os & 557 & $-$ & \cite{NuclPhysA.237.333} \\
$^{189}$Os & $-$ & 31  & \cite{PhysRevC.14.2095} \\
$^{191}$Os & $-$ & 0   & \cite{JPhysG.31.S1891} \\
$^{193}$Os & $-$ & 316 & present work \\
\bottomrule
\end{tabular*}
\vspace{0mm}
\end{center}

\section{SUMMARY}
The neutron rich nucleus $^{193}$Os was produced in the single neutron transfer reaction $^{192}$Os($^{7}$Li,$^{6}$Li)$^{193}$Os.
An isomeric state in $^{193}$Os was observed at the excitation energy of 315.6 keV by means of $\gamma$-$\gamma$ prompt and delayed coincidences.
Moreover, the level scheme connecting the isomeric state with known low-spin states was established.
The configuration of this isomeric state was assigned as $9/2^-$[505] by means of the hindrance factor $f_\nu$ as well as the level structure systematics in the Os isotopes.
This is the first observation of high-spin states in the neutron rich nucleus $^{193}$Os. The present work extended our knowledge of high-spin states in Os isotopes.

\end{multicols}

\begin{multicols}{2}

\acknowledgments{The authors are grateful to the staff of the inbeam $\gamma$-ray group and the tandem accelerator group at CIAE for their help.}

\end{multicols}

\begin{multicols}{2}

\end{multicols}

\clearpage

\end{document}